\documentclass[aps,prl,twocolumn,superscriptaddress,showpacs]{revtex4}

\usepackage{graphicx}
\usepackage{bm}
\usepackage{amssymb}

\begin{document}

\title{Coriolis Effect in Optics: Unified Geometric Phase and Spin-Hall Effect}

\author{Konstantin Y. Bliokh}

\affiliation{Micro and Nanooptics Laboratory, Faculty of Mechanical Engineering, and
Russel Berrie Nanotechnology Institute, Technion--Israel Institute of Technology, Haifa
32000, Israel}

\affiliation{Institute of Radio Astronomy, 4 Krasnoznamyonnaya st., Kharkov 61002,
Ukraine }

\affiliation{Nonlinear Physics Center, Research School of Physical Sciences and
Engineering, Australian National University, Canberra ACT 0200, Australia}

\author{Yuri Gorodetski}

\affiliation{Micro and Nanooptics Laboratory, Faculty of Mechanical Engineering, and
Russel Berrie Nanotechnology Institute, Technion--Israel Institute of Technology, Haifa
32000, Israel}

\author{Vladimir Kleiner}

\affiliation{Micro and Nanooptics Laboratory, Faculty of Mechanical Engineering, and
Russel Berrie Nanotechnology Institute, Technion--Israel Institute of Technology, Haifa
32000, Israel}

\author{Erez Hasman}

\affiliation{Micro and Nanooptics Laboratory, Faculty of Mechanical Engineering, and
Russel Berrie Nanotechnology Institute, Technion--Israel Institute of Technology, Haifa
32000, Israel}

\begin{abstract}
We examine the spin-orbit coupling effects that appear when a wave carrying intrinsic
angular momentum interacts with a medium. The Berry phase is shown to be a manifestation
of the Coriolis effect in a non-inertial reference frame attached to the wave. In the
most general case, when both the direction of propagation and the state of the wave are
varied, the phase is given by a simple expression that unifies the spin redirection Berry
phase and the Pancharatnam--Berry phase. The theory is supported by the experiment
demonstrating the spin-orbit coupling of electromagnetic waves via a surface plasmon
nano-structure. The measurements verify the unified geometric phase, demonstrated by the
observed polarization-dependent shift (spin-Hall effect) of the waves.
\end{abstract}

 \maketitle

\emph{Introduction.---} Geometric phase is an inherent feature of the polarization optics
that appears under evolution of electromagnetic waves in inhomogeneous and anisotropic
media \cite{1}. There are two types of geometric phases: (i) the spin redirection
Rytov--Vladimirskii--Berry phase associated with the parallel transport of the wave field
under SO(3) variations of the direction of propagation of the wave \cite{2,3} and (ii) the
Pancharatnam--Berry phase that occurs under SU(2) manipulations with the polarization
state of light \cite{4,5}. In the general case, when both the direction of propagation and
polarization state of the wave are varied, the geometric phase becomes more intricate
\cite{6,7,8,9}, and a unified geometrical description of the phase requires the tricky Majorana
representation \cite{8}.

Another generalization of the geometric phases appears when one considers higher-order
wave beams carrying an intrinsic orbital angular momentum (AM) accompanied by a phase vortex \cite{10}. In
contrast to spin AM (polarization), which equals $\sigma=1$, the orbital AM may take
arbitrary integer values $\ell=0,1,2,...$ (we determine AM per one photon in the units of
$\hbar$). The geometric and dynamic properties of light beams bearing orbital AM resemble
behavior of massless particles with spin $j=\sigma+\ell$. In particular, the spin
redirection Berry phase and the Pancharatnam--Berry phase are generalized to the states
of light with $\ell\neq 0$ \cite{11,12}. While the spin redirection phase is still associated
with the parallel transport under SO(3) evolution \cite{11}, the Pancharatnam-type phase
arises from SU($2j$) manipulation with the modes \cite{12,13}. Accordingly, the Majorana
formalism becomes drastically complicated for higher spins \cite{14}.

Whereas the Majorana representation develops the geometrical interpretation, the Berry
phase can also be treated dynamically, as a manifestation of the spin-orbit interaction
of waves carrying AM \cite{9,11,15,16}. (In using the term ``spin-orbit interaction'' we refer
to the interaction between intrinsic AM and the extrinsic evolution of the wave.) In such
approach, the geometric phase turns out to be closely related to the Coriolis effect
\cite{17}. Furthermore, it has recently been shown that the spin redirection geometric phase
is accompanied by the spin-Hall effect of light \cite{9,11,15,16,18}, i.e., an AM-dependent
displacement of the wave trajectory, which argues in favor of the dynamical
interpretation of the Berry phase.

In this Letter, we develop and unify the dynamical approach to the geometric phases in
optics. We show that in the most general case, when a wave carrying an arbitrary AM
changes its direction of propagation and state, the geometric phase is still given by a
rather simple expression stemming from the Coriolis effect. The theoretical conclusions
are confirmed experimentally using a focusing surface plasmon nano-structure. We observe
that the geometric phases of partial waves bring about a polarization-dependent shift of
the resulting intensity distribution which sheds light on the nature of the spin-Hall
effect and explains other recent experiments. Our formalism can be equally applied to a
number of problems outside the optical field. Indeed, the Coriolis effect is responsible
for the Berry phase in classical \cite{19} and quantum \cite{20} systems, and the spin-Hall effect
occurs in the evolution of various quantum particles \cite{16,21}.

\emph{Coriolis effect.---} The total intrinsic AM of a paraxial wave equals ${\bf J} =
J{\bf u}$, where ${\bf u} = {\bf k}/k$ is the vector indicating the direction of
propagation of the wave (${\bf k}$ is the wave vector), and $J$ is the mean helicity of
the wave which has a continuous spectrum in the range $\left[ {- j,j} \right]$.

We consider a coherent non-collinear bundle of waves which are either (A) emitted or (B)
scattered by a motionless body, Fig.~1. Let the observer measures the wave
characteristics in a coordinate frame moving across the bundle and, hence, rotating with
some instantaneous angular velocity ${\bf \Omega}$, Fig.~1. Although the system is
stationary in the laboratory reference frame, the waves successively measured by the
observer are related by the precession equation: $d{\bf u}/dt = {\bf \Omega } \times {\bf u}$.
The frequencies of the waves measured in the rotating frame suffer the shift (case A)
\cite{22}
\begin{equation}\label{eq1}
\Delta \omega  =  - {\bf J} \cdot {\bf \Omega }~.
\end{equation}
This shift corresponds to an analogous term in the wave Hamiltonian which is responsible
for the spin-orbit interaction.  In the case (B), the observer will register the
frequency difference between the input and output waves:
\begin{equation}\label{eq2}
\delta\omega = \Delta\omega - \Delta\omega_0 =  - ({\bf J}-{\bf J}_0) \cdot
{\bf \Omega }~,
\end{equation}
where the subscript 0 indicates the characteristics of the input wave, Fig.~1b.

The frequency shift (\ref{eq1}) is a manifestation of the Coriolis effect in the non-inertial
reference frame. It is identical to the frequency shift of the rotational modes of the
Foucault pendulum staying on the rotating Earth ($\bf J$ is orthogonal to the Earth's
surface, and the angle $\theta$ between $\bf J$ and $\bf \Omega$ is determined by the
latitude $\pi/2-\theta$, Fig. 1) \cite{19}. The shifts (\ref{eq1}) and (\ref{eq2}) are similar to the angular
Doppler shifts (see in \cite{10}) that are produced as the medium moves with respect to
observer with the angular velocity $- \bf \Omega$. However, unlike the real angular
Doppler shift, there is \textit{no} real rotation of the medium and energy exchange in our case.
The Coriolis shifts (\ref{eq1}) and (\ref{eq2}) are artefacts of the rotation of the reference frame, and they
indicate the \textit{inertia} of the wave field.

\emph{Unified geometric phase.---} The Berry phase for waves evolving in stationary media
is a manifestation of the Coriolis effect. Indeed, the non-inertial reference frame
appears when the wave phase is measured in the coordinate frame attached to the varying
direction of propagation $\bf u$ \cite{17} (the case of the spin redirection phase) or to the
polarization ellipse (the case of the Pancharatnam phase). In both cases, the geometric
phase $\Phi = \int {\Delta \omega} \,dt$ originates from the Coriolis shift (\ref{eq1}):
\begin{equation}\label{eq3}
\Phi = - \int {{\bf J} \cdot {\bf \Omega }} \,dt~.
\end{equation}
This phase is added to the wave phase calculated in a parallel-transport (i.e., with ${\bf \Omega}\cdot{\bf J}=0$) reference frame (see Ref.~10c).
Although we considered the temporal evolution of the coordinate frame, Eq. (\ref{eq3}) holds true
for spatial evolution along a coordinate $\zeta$ as well. This case is described via the
substitution
\begin{equation}\label{eq4}
dt \to d\zeta~,~~{\bf \Omega } \to {\bf \Omega }_\zeta~,
\end{equation}
where ${\bf \Omega }_\zeta$ is the rate of rotation of the coordinate frame with respect
to the $\zeta$ coordinate.

\begin{figure}[t]
\centering \scalebox{0.37}{\includegraphics{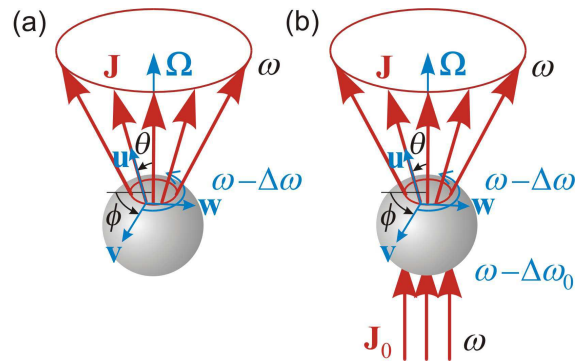}} \caption{(Color online.) The
Coriolis effect observed in the rotating reference frame $({\bf u},{\bf v},{\bf w})$.}
\label{Fig1}
\end{figure}

First, when the wave in an AM eigenstate with integer $J={\rm const}$ propagates along a curved trajectory (so that
the direction $\bf u$ is varied), and $\zeta$ is the coordinate along this trajectory,
then the phase $\Phi$ is the spin redirection Berry phase \cite{17}. If $\bf u$ is given by
the spherical angles $\left( {\theta ,\phi } \right)$ in the laboratory frame, then the
attached reference frame $\left( {{\bf v},{\bf w},{\bf u}} \right)$, Fig. 1, suffers
rotation with $\Omega _\zeta   = d\phi /d\zeta$ and ${\bf J} \cdot {\bf \Omega }_\zeta =
J\cos \theta {d\phi }/{d\zeta }$. Then, Eqs. (\ref{eq3}) and (\ref{eq4}) result in
\begin{equation}\label{eq5}
\Phi  =  - J\int {\cos \theta } \,d\phi~,
\end{equation}
which is the known expression for the spin redirection Berry phase \cite{1,3,17,23}.

Second, let the wave now propagates along the $\zeta$-coordinate in a fixed direction
${\bf u}={\rm const}$, but the helicity $J$ is continuously varied. This is accompanied by mixing of AM eigenstates and appearance of an oriented transverse structure of the wave (e.g., polarization ellipse). In the simplest case of $j=1$, the
state of the wave can be represented on the Poincar\'{e} (Bloch) sphere. In the spherical
coordinates $\left( {\vartheta ,\varphi } \right)$ on the Poincar\'{e} sphere, the wave
helicity equals $J=\cos\vartheta$, whereas the angle of rotation of the polarization
ellipse is $\varphi/2$. The wave-accompanying coordinate frame $\left( {{\bf v},{\bf w},{\bf u}} \right)$ with ${\bf v}$ attached to the polarization ellipse rotates with the angular velocity $\Omega _\zeta   = \left( {1/2} \right)d\varphi /d\zeta$, and
Eqs. (\ref{eq3}) and (\ref{eq4}) yield
\begin{equation}\label{eq6}
\Phi  =  - \frac{1}{2}\int {\cos \vartheta } \,d\varphi~,
\end{equation}
which is the known expression for the Pancharatnam--Berry phase \cite{1,5,23}. For higher
angular momenta, $j>1$, the wave state is described by ${\rm SU}\left( {2j} \right)$
structure \cite{12,13}, and cannot be represented on the Poincar\'{e} sphere.

Finally, in the most general case of continuous evolution of a wave carrying an AM $j$,
with both the direction and absolute value of ${\bf J}$ being varied, the higher-spin
Majorana representation can be evoked. This advanced geometrical formalism represents the
wave state by a set of $2j$ vectors ${\bf n}_i$, $i=1,...,2j$, on a unit sphere
generalizing the $\bf u$ sphere of directions \cite{14}. The geometric phase for cyclic
evolutions is expressed by a rather complicated contour integral involving motion of
these vectors on the sphere. Surprisingly, substitution of the rotational evolution
${d{\bf n}_i}/{d\zeta} = {\bf \Omega }_\zeta \times {\bf n}_i$ reduces this
integral to the simple expression (\ref{eq3}) for $\Phi$ (see Eq.~(20) in \cite{14}). This expression
remains valid for non-cyclical evolutions provided the measurements are made in the
reference frame attached to the direction of propagation and the transverse structure of the field.

Similarly, the phase difference due to Eq.~(\ref{eq2}),
\begin{equation}\label{eq7}
\delta\Phi = - \int {\left( {{\bf J} - {\bf J}_0 } \right) \cdot {\bf \Omega }} \,dt~,
\end{equation}
determines the geometric phases in systems with an abrupt transformation of the AM from
${\bf J}_0$ to ${\bf J}$ and a continuous evolution of ${\bf J}$. In this case, we imply that the incident wave is in an AM eigenstate and ${\bf J}_0 \parallel {\bf \Omega}$. Otherwise, an additional geometric phase may appear due to motion of the $\bf J$-accompanying frame with respect to the the incident wave (see below).

\emph{Experiment: Geometric phase producing the spin-Hall effect.---} The experiment was
performed utilizing surface plasmon-polariton waves and their strong polarization
coupling sensitivity together with high surface confinement. The metallic sample, Fig. 2, has a semi-circular corrugation (the inner radius is 1640 nm) that acts as a
lens for the surface plasmons ($\lambda = 493$ nm) \cite{24}. The grating was etched by a
focused ion beam in a 100-nm-thick gold film evaporated onto a glass wafer. The coupling
grating (period of 500 nm) was accompanied by a Bragg grating (period of 250 nm) from the
outside. The grating grooves enclosed an azimuthal angle $\phi$ from 5 to 175 degrees in
order to avoid resonant edge effects. The sample was illuminated from the bottom by a
green laser ("Verdy" doubled Nd:Yag, $\lambda_0 =532$ nm), while the
near-field intensity of the surface plasmons was measured by a Near-field Scanning
Optical Microscope (NSOM; Nanonics Multiview 2000) in a non-contact mode.

The scheme of the experiment involving wave vectors and polarizations is shown in Fig. 2.
The incident light, propagating along the $z$ axis, is right- or left-hand circularly
polarized, ${\bf J}_0  =  \pm {\bf e}_z$ (AM eigenstates). The coupling grating
on the surface of the sample transforms the propagating waves into surface plasmons. They
are linearly polarized (the electric field is normal to the metal surface), so that ${\bf
J} = 0$, and have the $\bf k$ vectors directed normally to the grating in the $(x,y)$
plane. Neither the spin redirection phase nor the Pancharatnam phase is applicable in
this system, since the wave changes both its direction of propagation and polarization
state. At the same time, the phase can be readily calculated using Eq. (\ref{eq7}) with Eq. (\ref{eq4}).
The plasmon-accompanying coordinate frame $({\bf v},{\bf u},{\bf w})$ is locally
attached to the grating with ${\bf u}={\bf k}/k$ and ${\bf v}={\bf e}_z$. The azimuthal
angle along the grating equals to the azimuthal angle of the plasmon wave vector and can
be taken as the coordinate $\zeta=\phi$ in the above formalism. In this manner, ${\bf
\Omega }_\phi = {\bf e}_z$, and the phase difference appearing at the grating equals
\begin{equation}\label{eq8}
\delta\Phi  = \int {J_0 } \,d\phi  =  \pm \phi~.
\end{equation}

\begin{figure}[t]
\centering \scalebox{0.34}{\includegraphics{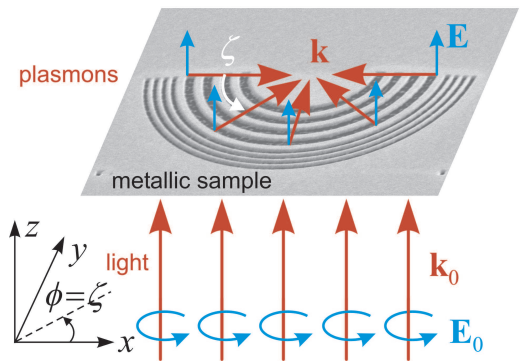}} \caption{(Color online.)
Scheme of the experiment.} \label{Fig2}
\end{figure}

Geometric phase (\ref{eq8}) describes an azimuthal phase gradient in the plasmon field that
depends on the polarization of the incident light. This changes the wave phase front into
a segment of spiral with the pitch $\lambda$ and the helicity dependent on the
polarization of the incident wave, Fig. 3. As the surface plasmons are focused by the
semi-circular geometry of the sample, the phase (\ref{eq8}) modifying the phase front brings
about the transverse shift of the resulting focal spot. This shift can be estimated as
\begin{equation}\label{eq9}
\delta x \sim \pm \frac{\lambda}{2\pi}~.
\end{equation}
Since the shift (\ref{eq9}) is proportional to the helicity of the incident wave, $J_0=\pm 1$,
it should be regarded as a spin-Hall effect of electromagnetic waves \cite{9,11,15,16,18}. The
spin-Hall effect originates from the polarization-dependent geometric phases (\ref{eq8}) of
partial waves forming the confined field in the focus. The polarization-dependent shift
(\ref{eq9}) is clearly seen both in experimental measurements and in numerical calculations
processing the evolution of field with the phase (\ref{eq8}), Fig. 3, thereby, confirming the
geometric phase in the system.

Being a superposition of the right- and left-hand circular modes, the linearly
$x$-polarized incident wave (which is not an AM eigenstate) brings about a split double
spot in the output plasmon field, Fig. 3, cf. \cite{24}. Such a field distribution is similar
to a ${\rm HG}_{1,0}$ beam and has a phase singularity: a $\pi$ jump at $x=0$, Fig. 3.
This phase jump is a Panchratnam-type phase that arises due to a $\pi$ rotation of the
polarization ellipse of the incident light with respect to the coordinate frame attached
to the grating.

\begin{figure}[t]
\centering \scalebox{0.33}{\includegraphics{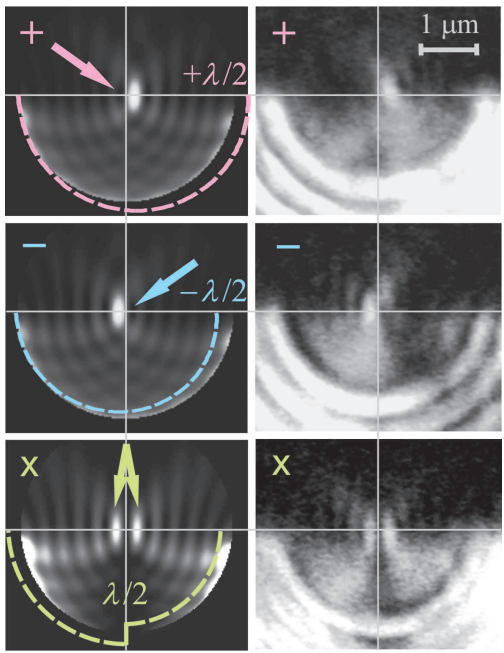}} \caption{(Color online.)
Intensities of the plasmon field in the sample for the right-hand (+), left-hand (--),
and linear $x$-polarizations (x) of the incident light. Left panels: numerical
calculations of the wave propagation with the phase fronts (dashed lines) disturbed by
the geometric phase (\ref{eq8}). Right panels: experimental measurements.} \label{Fig3}
\end{figure}

\emph{Discussion.---} We have shown that the geometric phase is unified in terms of
spin-orbit interaction and Coriolis effect and is given
by a simple expression. Our experiment has verified the unified geometric phase in a
rather complex system including light and surface plasmon-polaritons. The results reveal
the nature of the spin-Hall effect -- the spin-dependent geometric phase of partial waves
brings about a spin-dependent transverse shift of the resulting confined wave field. In
contrast to semiclassical theories implying paraxial confined waves \cite{15,16,18,21}, our
experiment dealt with a highly non-paraxial field. As a result, the transverse shift is
comparable with the focal spot width and is readily detected.

Polarization-dependent displacements of light beams diffracted by a space-variant grating
producing the Pancharatnam--Berry phase have been observed in \cite{25}. A
polarization-dependent transverse shift of the focal spot in lenses has also been
reported in papers \cite{26,27}. In \cite{26} this effect was explained in the context of the
spin-orbit interaction of photons in a vacuum. In a recent paper \cite{27}, deformation and
shifts of the wave intensity in the focus of a high-numerical-aperture lens were treated
as manifestations of the Pancharatnam--Berry phase and have been associated with the spin-Hall effect of light. Our results can be applied
to the systems in \cite{24,25,26,27}, clarifying the interrelations between the spin-orbit
interaction, geometric phase, and spin-Hall effect of light.

The work by K.B. is supported by the Linkage International Grant of the Australian Research Council.


\begin{references}
\bibitem{23}{The phases (\ref{eq5}) and (\ref{eq6}) are determined in the coordinate frame rotated with azimuthal
angles $\phi$ and $\varphi/2$, respectively. Subtraction of this rotation leads to the
substitution $- \cos \theta  \to 1 - \cos \theta$ and $- \cos \vartheta  \to 1 - \cos
\vartheta$, and integrals in Eqs. (\ref{eq5}) and (\ref{eq6}) give the solid angles enclosed by the wave
evolution on the corresponding spheres \cite{1,2,3,4,5}.}
\end{references}
\end{document}